\documentclass[sigconf]{acmart}

\copyrightyear{2026}
\acmYear{2026}
\setcopyright{cc}
\setcctype{by}
\acmConference[UbiComp Companion '26]{Companion of the 2026 ACM International Joint Conference on Pervasive and Ubiquitous Computing}{October 11--15, 2026}{Shanghai, China}
\acmBooktitle{Companion of the 2026 ACM International Joint Conference on Pervasive and Ubiquitous Computing (UbiComp Companion '26), October 11--15, 2026, Shanghai, China}
\acmDOI{10.1145/3798063.3841767}
\acmISBN{979-8-4007-2533-3/2026/10}
\microtypesetup{expansion=false}
\usepackage{booktabs}
\usepackage{graphicx}
\usepackage{array}
\usepackage{float}
\usepackage{tikz}
\usetikzlibrary{arrows.meta,positioning,fit,backgrounds,calc}


\begin{document}

\title{Oto-Meal: Earable Sensing with PPG and IMU for Personalized Meal Awareness}

\author{Yuxuan Hou}
\orcid{0009-0003-1938-0672}
\affiliation{%
  \institution{Southern University of Science and Technology}
  \city{Shenzhen}
  \state{Guangdong}
  \country{China}
}
\email{12413104@mail.sustech.edu.cn}

\author{Jiao Li}
\orcid{0000-0002-3918-4922}
\affiliation{%
  \institution{Shenzhen Polytechnic University}
  \city{Shenzhen}
  \state{Guangdong}
  \country{China}
}
\email{jiaoli@szpu.edu.cn}

\author{Linshan Jiang}
\orcid{0000-0001-8501-9488}
\affiliation{%
  \institution{Southern University of Science and Technology}
  \city{Shenzhen}
  \state{Guangdong}
  \country{China}
}
\email{jiangls@sustech.edu.cn}

\author{Jin Zhang}
\authornote{Corresponding author.}
\orcid{0000-0002-2674-0918}
\affiliation{%
  \institution{Southern University of Science and Technology}
  \city{Shenzhen}
  \state{Guangdong}
  \country{China}
}
\email{zhangj4@sustech.edu.cn}

\renewcommand{\shortauthors}{Yuxuan Hou, Jiao Li, Linshan Jiang, and Jin Zhang}

\begin{abstract}
Meal awareness can help people reflect on hydration, chewing rhythm, and conversation-heavy meals, but many eating-sensing approaches rely on cameras, microphones, food photographs, or repeated self-logging. PPG and IMU offer a narrower sensing path by capturing physiological and motion patterns around meal-adjacent actions without raw audio, video, or photographs. We present Oto-Meal, an audio- and image-free earable prototype. Its pooled neural recognizer uses a two-stage event/rest gate and five-class behavior classifier. Separately, a within-user protocol evaluates a lightweight memory matcher built from labeled target-user examples. We invited seven volunteers and collected a seven-user dataset for mixed-user training, within-user memory evaluation, and modality ablation. The pooled model reaches 70.99\% event accuracy. Under the separate memory protocol, 20\% target-user calibration reaches 80.38 $\pm$ 0.84\% event accuracy and 81.77 $\pm$ 0.69\% cascade accuracy; with 60\% calibration, PPG+IMU reaches 85.13 $\pm$ 0.57\% event accuracy and outperforms IMU-only and PPG-only. These preliminary results suggest that audio- and image-free earable sensing with inspectable personalization can support low-burden meal-awareness review.
\end{abstract}



\maketitle

\begin{figure*}[t]
\centering
\includegraphics[width=\textwidth]{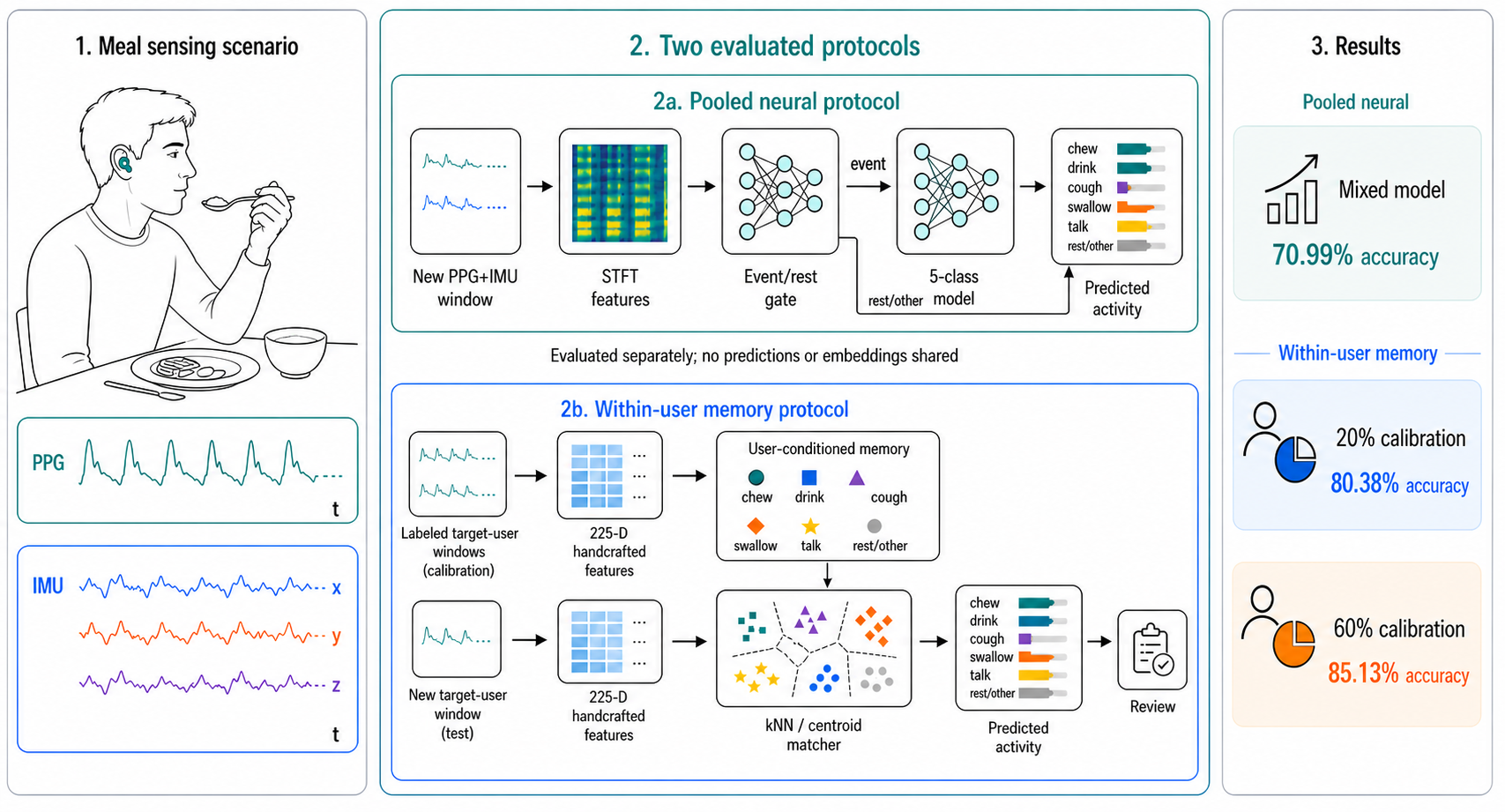}
\caption{System overview of Oto-Meal. The left panel shows the meal-sensing scenario and PPG+IMU input streams. The center panel presents two separately evaluated protocols: the pooled neural protocol maps STFT features through event/rest and five-class models, whereas the within-user protocol maps labeled and test target-user windows to 225-dimensional handcrafted features and a kNN/centroid matcher. The protocols share neither predictions nor embeddings. The right panel summarizes representative seven-user event-accuracy operating points; exact metrics appear in Figure~\ref{fig:overall-accuracy} and Table~\ref{tab:main}.}
\Description{A wide system overview. The left panel shows a person wearing an earable and the recorded PPG and three-axis IMU signals. The upper center lane shows the pooled neural protocol from a new PPG and IMU window through STFT features, an event-rest gate, and a five-class model. The lower center lane shows the separate within-user memory protocol: labeled calibration and test windows become 225-dimensional handcrafted features, and a kNN or centroid matcher predicts one of five activities or rest. The right panel lists mixed-model, 20-percent calibration, and 60-percent calibration event-accuracy operating points.}
\label{fig:architecture}
\end{figure*}

\section{Introduction}

Meals are recurring moments where well-being, routine, and social context intersect. They shape hydration, chewing rhythm, conversation, stress routines, and small daily choices that people may want to review later. Unlike step counting or sleep duration, however, meal behavior happens in socially dense settings: a meal can include companions, speech, locations, food preferences, and repeated routines. A meal-awareness system therefore has to support reflection without turning the table into a high-burden logging task or a broad context recorder.

Prior work~\cite{bi2016autodietary,choe2014understanding,li2010stage,luo2019codesigning,lu2022semiautomated,meyers2015im2calories} highlights the difficulty of meal-awareness sensing under privacy and burden constraints. Manual food diaries and personal informatics tools can support reflection, but they require repeated user effort and careful integration into everyday routines~\cite{li2010stage,choe2014understanding,luo2019codesigning}. Semi-automated dietary systems reduce some effort, yet users still need to know what is sensed, what is inferred, and how uncertain outputs should be interpreted~\cite{lu2022semiautomated}. Vision-based food journaling can reason about food content~\cite{meyers2015im2calories}, while acoustic systems can detect intake sounds~\cite{bi2016autodietary}. These routes are useful, but raw images and audio may capture companions, conversations, and food identity beyond the behavior-level reflection that a user actually wants.

This paper therefore asks a narrower sensing question: can an earable-style PPG+IMU pipeline provide reviewable meal-adjacent behavior timelines without raw audio, video, or food photos? Recent earable works~\cite{ketmalasiri2024imchew,li2023earpass,montanari2023earset} show that earphone IMUs can support chewing analysis and that in-ear PPG carries useful biometric and motion-sensitive information. Oto-Meal builds from this sensing direction but targets a broader behavior timeline: chewing, drinking, coughing, swallowing, talking, and rest/other. These labels are low-burden cues for self-awareness, not nutrient estimates or clinical judgments.

However, the main technical challenge is that physiological and inertial signals vary across users, sessions, sensor placement, and behavior style. PPG and IMU windows can shift with physiology, earbud fit, head motion, and the way a person performs short actions such as drinking or swallowing. A pooled model may learn shared structure, but meal-awareness review also needs a practical way to adapt to the target user's signal style.

Oto-Meal addresses this challenge with a two-part design evaluated under separate protocols. First, a pooled PPG+IMU recognizer converts windows into an event/rest decision and a five-class behavior label. Second, under a separate within-user protocol, a lightweight user-conditioned memory uses a small labeled calibration set to store target-user examples and perform memory-based matching with handcrafted PPG+IMU features. The matcher does not consume the pooled model's predictions or embeddings and does not retrain that model. The memory layer is deliberately simple: its stored calibration examples remain directly inspectable, and it is evaluated as a calibration component rather than hidden inside a full retraining pipeline. In the current prototype, sensing data are processed in a laptop workflow; we treat this as a feasibility step toward future lightweight or on-device implementations, not as a completed field deployment.

To demonstrate the benefit of this design for meal-awareness recognition, we conduct experiments under the evidence available in our seven-user dataset. The experiments cover mixed-user training, target-user calibration, modality ablation, and runtime measurement. The main metrics are event accuracy for the five behavior classes and cascade accuracy for the full event/rest plus behavior-label decision. The evaluation reports two separate operating settings. The mixed-user model is useful, while event and cascade accuracy under the target-user memory protocol cross 80\% with 20\% labeled calibration and continue improving with larger calibration ratios. Because the settings use different representations and classifiers, they are not treated as a controlled ablation of personalization.

In summary, our main contributions can be summarized as follows:
\begin{itemize}
    \item To avoid rich contextual media collection, we frame audio- and image-free PPG+IMU sensing as a less intrusive primitive for meal-awareness timelines.
    \item To make personalization bounded and inspectable, we describe a two-stage recognizer plus user-conditioned memory that separates preparation and calibration from inference and use.
    \item To demonstrate the effectiveness of our approach, we report our seven-user evidence covering mixed-user training, target-user calibration, modality ablation, and runtime.
\end{itemize}

\section{Related Work}

\subsection{Meal-awareness scenarios and self-tracking}

Meal awareness asks how sensing can help users review eating-adjacent routines without turning meals into a heavy logging task. Personal informatics research shows that self-tracking systems fail when they treat sensing as the whole problem. Users must still collect, integrate, reflect on, and act on personal data~\cite{li2010stage}, and quantified-self practices often depend on how data can be explored and made meaningful later~\cite{choe2014understanding}. Food tracking makes this burden concrete. Dietitians and users need records that are customizable and interpretable rather than generic logs~\cite{luo2019codesigning}. Semi-automated dietary monitoring further shows that sensing capability and privacy tradeoffs must be communicated clearly~\cite{lu2022semiautomated}. Oto-Meal builds on this scenario by treating recognition as one component of a reviewable meal-awareness timeline, not as automatic dietary authority.

\subsection{Earable and wearable PPG+IMU sensing}

Earable and wearable sensing has become an important route for eating-related behavior recognition, especially when systems use inertial, earable, or physiological signals instead of only self-report. Wrist IMU methods recognize eating moments with watches~\cite{thomaz2015practical}. Earphone and ear-adjacent methods are especially relevant: IMChew studies chewing analysis with earphone IMUs~\cite{ketmalasiri2024imchew}, while EarBit~\cite{bedri2017earbit}, Auracle~\cite{bi2018auracle}, and NeckSense~\cite{zhang2020necksense} show alternative body locations and sensor combinations for eating-related detection. PPG-based work further shows that the ear is a meaningful physiological sensing site: EarPass uses in-ear PPG for continuous authentication~\cite{li2023earpass}, EarSet characterizes how head and facial movement affect in-ear PPG~\cite{montanari2023earset}, EarFusion studies in-ear audio+PPG fusion for heart-rate monitoring~\cite{liu2025earfusion}, and recent earable work explores continuous blood-pressure monitoring with a single-point flexible sensor~\cite{li2025earablebp}. Clinical swallowing work also demonstrates the health relevance of wearable swallowing signals~\cite{shin2024automatic}. Oto-Meal also belongs to this earable and wearable sensing family, but its paper-facing design takes a stricter input stance: it uses PPG and IMU only, avoids raw audio and food images, and evaluates multiple meal-adjacent behavior labels rather than only eating-episode detection or clinical swallowing assessment.

\subsection{Personalization for well-being sensing}

Personalization is important for low-burden well-being sensing. PPG is often associated with cardiovascular monitoring, but earphone PPG can also encode facial-expression-related changes~\cite{choi2022ppgface}. This suggests that PPG near the ear may carry behavior-relevant signals beyond heart rate, especially when paired with IMU motion cues. At the same time, wearable signals are shaped by physiology, placement, motion, and session context. Personalized earbud snacking detection~\cite{morshed2022personalized} and few-shot mobile sensing adaptation~\cite{gong2019metasense} motivate target-user adaptation when a single global model is not enough. Oto-Meal adopts this principle with a small memory-based calibration matcher. The contribution is not a new personalization theory; it is preliminary evidence that a transparent calibration memory can support audio- and image-free meal-awareness recognition in a small seven-user, within-user setting.

\section{System and Method}

\subsection{Scenario and system architecture}

Oto-Meal targets reflective meal awareness for a user who wants to review behavior patterns after a meal period: for example, frequent drinking-like segments, extended talking, or uncertain drink/swallow regions. Figure~\ref{fig:architecture} summarizes the two-phase workflow. During preparation and calibration, synchronized PPG+IMU recordings are converted into normalized windows, and labeled examples are used to build a target-user memory. During inference and review, new PPG+IMU windows are preprocessed, compared with the user-conditioned memory, and converted into predicted activity labels for review. The two-stage recognizer that supplies the event/rest and event-class structure is described below. This separation matters because the user's everyday workflow should be inference and review, while labeling and calibration should remain bounded setup steps.

The modules in Figure~\ref{fig:architecture} map to the paper components at an overview level. The signal panels define what is sensed and what is deliberately excluded: PPG and IMU are kept, while raw audio, video, and food photos are not part of the pipeline. The preparation and inference modules summarize feature extraction, memory construction, memory-based matching, and activity prediction; the event/rest and behavior-label decision is detailed in the next subsection. The user-conditioned memory stores calibration examples as prototypes or nearest-neighbor entries. The final review output is therefore not a food diary with images; it is a sequence of behavior labels and uncertainty cues that can support lightweight reflection.

The system keeps low-level sensing evidence inside the recognition pipeline and presents users with a higher-level, reviewable meal timeline. Raw time-series windows are transformed into labels and confidence-like scores; the user-facing object is a compact timeline rather than a sensor trace. This distinction is important for well-being use: a person may want to know that a meal included several drinking-like segments or a long talking-heavy region, but they should not have to inspect PPG waveforms. The system figure therefore emphasizes preparation, inference, and review as separate responsibilities rather than a single opaque classifier.

\subsection{Two-stage recognizer and user memory}

The recognition task has two stages. A binary gate separates event-like windows from rest/other. Event-like windows are then classified into five behavior classes. For the pooled neural recognizer, the binary and five-class models are trained separately. At cascade inference, a binary rest prediction maps directly to rest/other; otherwise, the five-class model supplies the final event label. This structure follows the practical use case: a meal timeline first needs to know whether a window contains a meaningful action, and only then whether the action is chew, drink, cough, swallow, or talk. We report binary accuracy for the first stage, event accuracy for true event windows, and cascade accuracy for the complete event/rest plus event-class decision.

The user-conditioned memory is evaluated independently of the pooled model. For each target user, class-stratified calibration windows form a per-user memory and the remaining windows are held out. Each six-channel PPG+IMU window becomes a 225-dimensional vector containing robust and first-difference statistics, five spectral-band fractions, spectral centroid and entropy, and pairwise cross-channel correlations. Calibration-set median and interquartile range standardize the features, with standard deviation used for a degenerate interquartile range. Candidates comprise cosine and Euclidean class centroids, cosine kNN with $k\in\{1,3,5,7,9\}$, Euclidean 5-NN, and two-stage 5-NN/centroid variants. Centroids are arithmetic means; kNN uses majority voting with class-index tie breaking. For each seed, ratio, modality, and split unit, the reported candidate is selected post hoc by held-out event accuracy, then cascade and binary accuracy. The matcher outputs hard labels without calibrated confidence, uses no pooled-model predictions or embeddings, and does not retrain the pooled model. Thus, personalized results assume labeled target-user examples and are not zero-shot claims.

\subsection{User workflow}

A practical workflow has four steps: a user provides labeled calibration windows; Oto-Meal builds the per-user memory on a laptop-class workflow; new windows become behavior timelines; and a future interface could accept corrections. The study evaluates only the sensing and memory components. It does not measure elapsed calibration time, user effort, confidence calibration, correction interaction, or longitudinal use, and it does not claim field deployment. The intended interaction is post-meal timeline review rather than continuous classifier monitoring.

Calibration is framed as a bounded setup cost rather than an ongoing diary requirement. A user does not need to label every future meal; instead, the system uses labeled examples to establish how that user's chewing, drinking, swallowing, and talking patterns appear in the PPG+IMU feature space. Later corrections could be treated as memory maintenance in a future interface. This is why the evaluation reports multiple calibration ratios: the ratios are an experimental proxy for how much labeled target-user evidence the memory has before it is used for review. A calibration ratio $r$ assigns approximately $r$ of each target user's windows within every class to memory construction and uses the remaining windows for testing. The 20\% setting contains 1,405 calibration windows and 5,623 test windows in aggregate. The ratio measures labeled-window availability; it is not 20\% of future meals or a measurement of elapsed calibration time.

\section{Evaluation}

\subsection{Experiments settings and metrics}

We invited seven volunteers, ages 17--33, and collected the dataset summarized in Table~\ref{tab:dataset}: 538 raw text entries and 7,028 generated windows covering chew, drink, cough, swallow, talk, and rest/other. Counts are 1,230, 1,174, 1,234, 1,096, 1,246, and 1,048, respectively. User support ranges from 116 windows for each of two users to 1,782 for the largest user. We therefore interpret aggregate accuracy with per-class recall and per-user results (Figures~\ref{fig:class-recall} and~\ref{fig:per-user-cascade}).

The event configuration provides labels, nominal starts, and durations. Local activity refines starts within $\pm0.8$~s, with no pre/post expansion or sliding overlap; gap-generated rest uses full-window stride. The set contains 6,219 one-second, 776 two-second, 24 five-second, and 9 ten-second windows: 6,232 configured-interval windows (including 252 explicit rest/other) and 796 gap-generated rest windows using a 0.5~s guard and activity filtering. Processing assumes 50~Hz; physical device rate and synchronization require separate hardware documentation.

The pooled recognizer receives six-channel $6\times32\times8$ log-magnitude STFT tensors. PPG is first-differenced, IMU remains raw, and each channel window is robust-normalized. The STFT uses a 64-point FFT, 32-sample Hann window, 4-sample hop, 32 interpolated bins, eight frames, and training-only global standardization. Separate PPG/IMU convolutional front ends use 32/64/128 channels with Group Normalization and ReLU; a learned gate precedes a two-layer, four-head Transformer (128-dimensional embedding, 0.2 dropout). Binary and event models are trained separately with label-smoothed cross-entropy and a 0.4-weighted IMU auxiliary term in both logits and objective. Adam uses learning rate $3\times10^{-4}$, weight decay $10^{-2}$, batch size 64, six warmup epochs, cosine annealing, and patience-15 early stopping within 60 epochs. Binary/event training stops at epochs 29/56 and selects checkpoints 14/41. Random oversampling balances classes; augmentation applies independent 0.2-probability frequency/time masks and 0.05 modality dropout.

The memory path uses handcrafted PPG+IMU window features as detailed in Section~3.2. Features are standardized using calibration-set statistics before test windows are scored.

This preparation step keeps the paper's claims bounded. We do not assume semantic knowledge about food type, meal content, or conversation content. The recognizer only receives sensor windows and labels derived from the behavior protocol. This makes the evaluation narrower than full dietary monitoring, but it also makes the data-minimizing design claim clearer: Oto-Meal asks whether behavior-level meal awareness can be supported without collecting the richer contextual streams that many users may find intrusive.

We use three protocols. The mixed one-model baseline stratifies an 80/20 split within each user, pools training windows, trains one neural PPG+IMU model, and tests on held-out windows from the same users. The target-user protocol builds a separate handcrafted-feature matcher per user from 10\%, 20\%, 40\%, 60\%, or 80\% class-stratified calibration windows and evaluates the remainder. The modality ablation repeats the memory protocol with PPG+IMU, IMU-only, and PPG-only features. Because pooled and memory protocols use different representations and classifiers, they are distinct operating settings, not a controlled personalization effect.

We report binary accuracy for the event/rest gate, event accuracy for five-class prediction on event windows, and cascade accuracy for the complete decision. Cascade accuracy penalizes both event/rest and event-class errors. Given imbalanced class and user support, these aggregate metrics are accompanied by per-class recall and per-user cascade accuracy.

The calibration ratios should be read carefully. They are not claims that a final product should require a fixed percentage of all future meals to be labeled. Instead, they let us compare operating points under a controlled protocol: more labeled target-user windows should make the memory more representative, but the key design question is whether useful gains appear before calibration becomes burdensome.

\begin{table}[t]
\centering
\caption{Seven-user dataset used in this paper. Counts describe the anonymized, clean paper-facing configuration.}
\label{tab:dataset}
\small
\begin{tabular}{@{}lrl@{}}
\toprule
Property & Value & Note \\
\midrule
Users & 7 & complete target-label coverage \\
Age range & 17--33 & years \\
Raw entries & 538 & normalized text recordings \\
Windows & 7,028 & after slicing/rest generation \\
Event labels & 5 & chew, drink, cough, swallow, talk \\
Rest/other & 1 & background/non-event class \\
User range & 116--1,782 & windows per user \\
\bottomrule
\end{tabular}
\end{table}

\subsection{Overall performance}

\begin{figure}[t]
\centering
\includegraphics[width=0.92\linewidth]{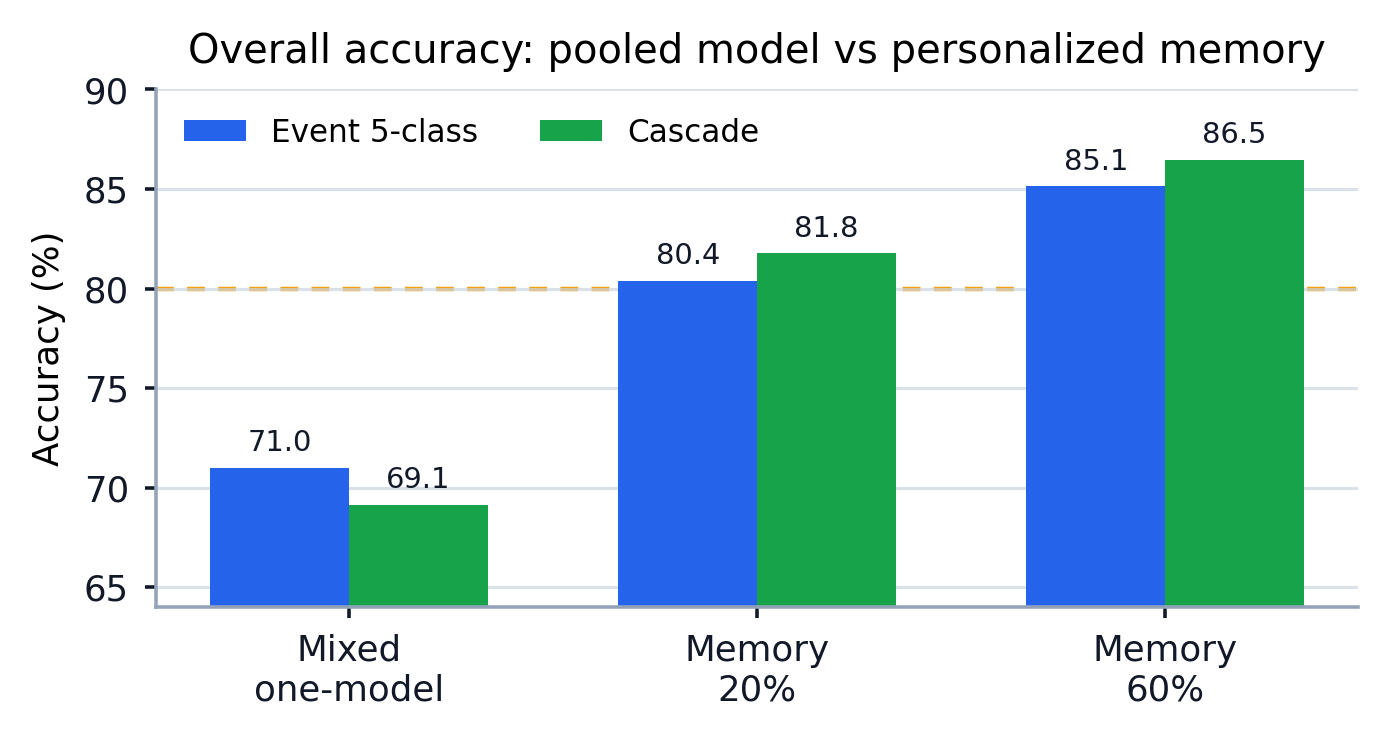}
\caption{Overall accuracy comparison across the pooled neural protocol and two within-user memory settings. The protocols use different representations and classifiers and are shown as distinct operating points.}
\Description{A bar chart comparing event and cascade accuracy for a mixed one-model baseline, 20 percent calibration memory, and 60 percent calibration memory.}
\label{fig:overall-accuracy}
\end{figure}

We first evaluate whether the audio- and image-free PPG+IMU path can support useful meal-awareness recognition when the target user's style is represented. Figure~\ref{fig:overall-accuracy} summarizes the main accuracy results across the two evaluation protocols. The pooled mixed-user model provides a useful operating point, reaching 70.99\% event accuracy and 69.11\% cascade accuracy. Under the separate within-user memory protocol, the 20\% setting reaches 80.38 $\pm$ 0.84\% event accuracy and 81.77 $\pm$ 0.69\% cascade accuracy. With 60\% calibration, PPG+IMU reaches 85.13 $\pm$ 0.57\% event accuracy and 86.47 $\pm$ 0.32\% cascade accuracy. These results are reported as distinct operating settings rather than as a controlled pooled-versus-personalization ablation.

\begin{figure}[t]
\centering
\includegraphics[width=0.92\linewidth]{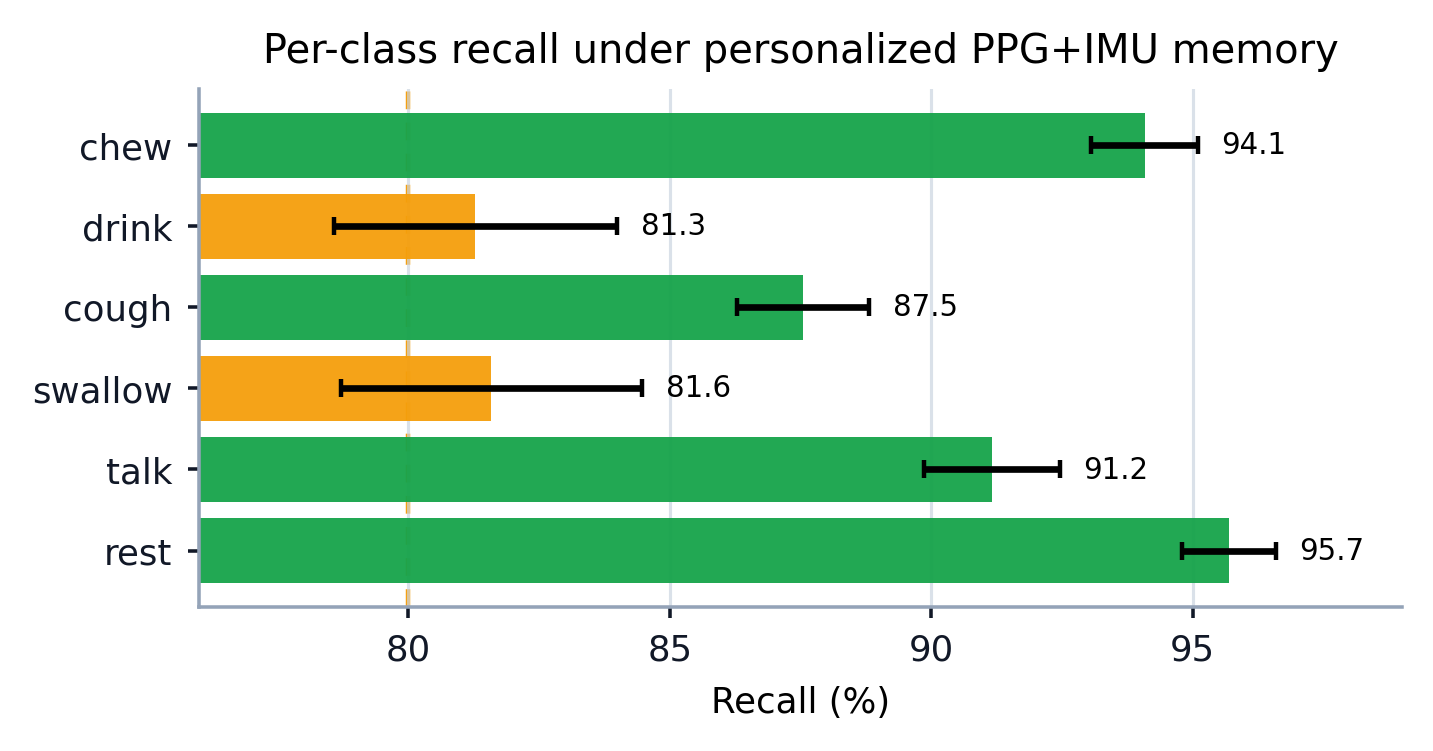}
\caption{Per-class recall under personalized PPG+IMU memory. Drink and swallow remain the closest event classes to the 80\% threshold.}
\Description{A horizontal bar chart showing recall for chew, drink, cough, swallow, talk, and rest. Drink and swallow are closest to the 80 percent threshold.}
\label{fig:class-recall}
\end{figure}

\begin{figure}[t]
\centering
\includegraphics[width=0.92\linewidth]{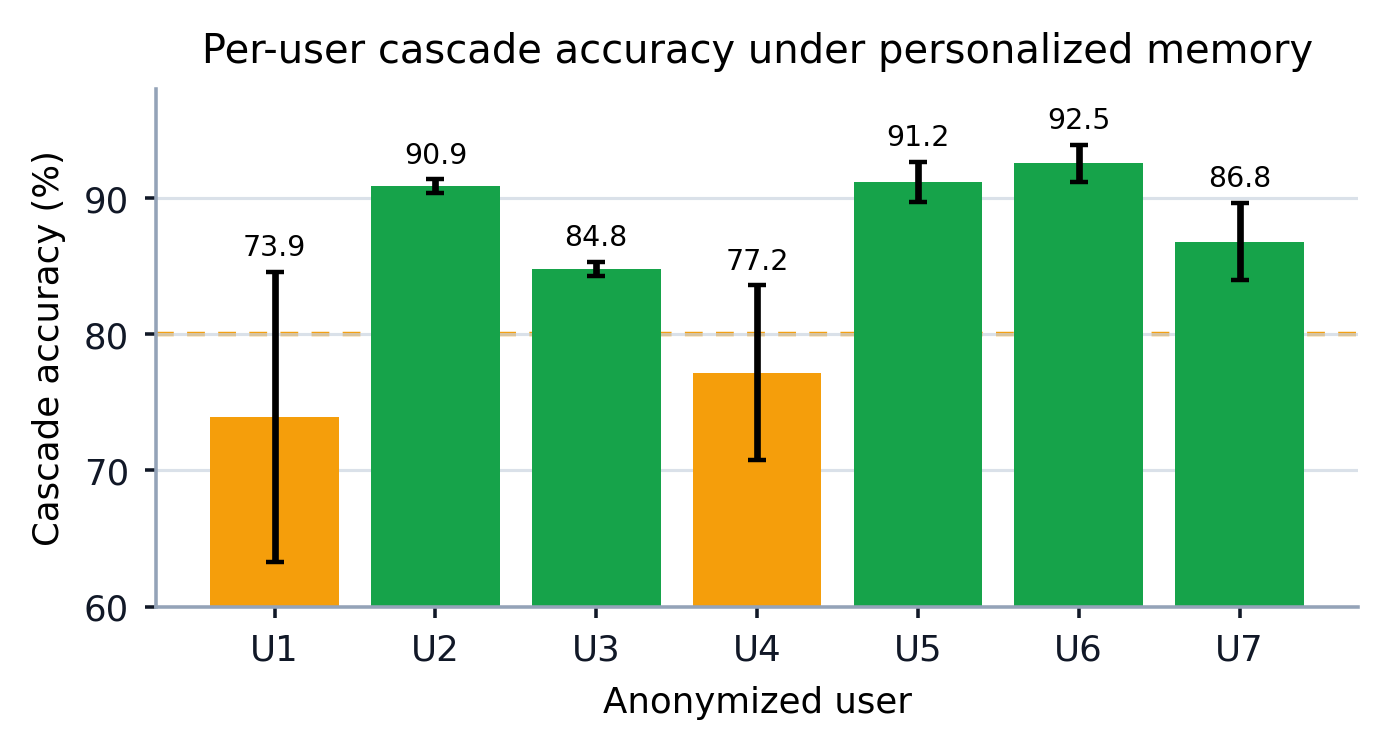}
\caption{Per-user cascade accuracy under personalized PPG+IMU memory. Most users exceed 80\% cascade accuracy, while smaller test sets have larger uncertainty.}
\Description{A bar chart showing cascade accuracy for seven anonymized users with error bars. Most users exceed 80 percent cascade accuracy.}
\label{fig:per-user-cascade}
\end{figure}

Figure~\ref{fig:class-recall} reports class-level recall under the stronger personalized PPG+IMU memory reference. Chew, talk, rest, and cough are relatively strong. Drink and swallow are closer to the 80\% recall threshold, matching the intuition that both involve short throat and head-motion patterns. Figure~\ref{fig:per-user-cascade} shows that per-user results also vary: most users exceed 80\% cascade accuracy, while estimates for users with fewer held-out windows are more volatile.

The memory path is also lightweight once windows are available. On a Windows laptop-class machine, a five-repeat timing pass over the existing 7,028 cached windows loaded and extracted features at 583.0 windows/s on average. For the 20\% calibration setting, building per-user memory and predicting 5,623 held-out windows took 40.5 ms on average, or about 138,831.7 test windows/s. These timings exclude neural-model training and should be interpreted as the cost of the lightweight personalized memory stage. They support the design choice of keeping personalization outside the deep model: memory updates are inexpensive enough to be triggered by corrected examples rather than by full retraining.

\begin{table}[t]
\centering
\caption{Seven-user result ladder. Calibration rows report four-seed mean $\pm$ standard deviation.}
\label{tab:main}
\footnotesize
\resizebox{\linewidth}{!}{%
\begin{tabular}{@{}lccc@{}}
\toprule
Protocol & Binary & Event & Cascade \\
\midrule
Mixed one-model & 92.38 & 70.99 & 69.11 \\
Memory, 10\% calib. & 94.94 $\pm$ 0.22 & 75.26 $\pm$ 1.60 & 76.90 $\pm$ 1.29 \\
Memory, 20\% calib. & 95.76 $\pm$ 0.20 & 80.38 $\pm$ 0.84 & 81.77 $\pm$ 0.69 \\
Memory, 40\% calib. & 96.38 $\pm$ 0.11 & 83.30 $\pm$ 0.15 & 84.50 $\pm$ 0.17 \\
Memory, 60\% calib. & 96.98 $\pm$ 0.31 & 85.13 $\pm$ 0.57 & 86.47 $\pm$ 0.32 \\
Memory, 80\% calib. & 97.26 $\pm$ 0.25 & 87.31 $\pm$ 1.05 & 88.56 $\pm$ 0.86 \\
\bottomrule
\end{tabular}
}
\end{table}

\subsection{Modality ablation}

Table~\ref{tab:modalities} summarizes the modality ablation under the same 60\% user-conditioned memory protocol. This ablation tests whether PPG adds value beyond motion alone, rather than only increasing sensor count. PPG+IMU outperforms either modality under the same memory protocol. At 60\% calibration, PPG+IMU is 4.81 percentage points higher than IMU-only in event accuracy and 15.31 points higher than PPG-only. The gap is largest for event and cascade accuracy, which are the metrics that matter most for a reviewable meal timeline.

The single-modality rows also clarify what each signal contributes. PPG-only keeps high binary accuracy, suggesting that it can help separate active meal-adjacent moments from rest, but its much lower event accuracy indicates that PPG alone does not provide enough class-discriminative structure for behaviors such as drink, swallow, talk, and chew. IMU-only is stronger because head, jaw, and throat motion are directly related to many target behaviors, but it still falls below the fused path. The fused result therefore supports the paper's sensing claim: PPG and IMU are complementary for behavior-level meal awareness. In practical terms, the PPG channel is not presented as a replacement for motion sensing; it is useful because it adds a physiological and contact-sensitive signal stream that improves the personalized memory when paired with IMU.

\begin{table}[t]
\centering
\caption{Modality ablation under the same 60\% user-conditioned memory protocol.}
\label{tab:modalities}
\footnotesize
\resizebox{\linewidth}{!}{%
\begin{tabular}{@{}lccc@{}}
\toprule
Modality & Binary & Event & Cascade \\
\midrule
\textbf{PPG+IMU} & \textbf{96.98 $\pm$ 0.31} & \textbf{85.13 $\pm$ 0.57} & \textbf{86.47 $\pm$ 0.32} \\
IMU-only & 95.18 $\pm$ 0.27 & 80.32 $\pm$ 0.77 & 81.43 $\pm$ 0.57 \\
PPG-only & 96.00 $\pm$ 0.25 & 69.82 $\pm$ 0.76 & 72.54 $\pm$ 0.66 \\
\bottomrule
\end{tabular}
}
\end{table}

\subsection{Performance under different calibration ratios}

Table~\ref{tab:main} reports how performance changes as the amount of labeled target-user calibration increases. In the separate target-user memory protocol, 10\% labeled calibration reaches 75.26 $\pm$ 1.60\% event accuracy and 76.90 $\pm$ 1.29\% cascade accuracy. The 20\% setting is the first memory operating point that crosses 80\% on both metrics, reaching 80.38 $\pm$ 0.84\% event accuracy and 81.77 $\pm$ 0.69\% cascade accuracy. The pooled model's 70.99\% event and 69.11\% cascade results are reported separately and are not a no-calibration condition for this matcher.

The later calibration points show a smoother refinement pattern. Moving from 20\% to 40\% calibration adds nearly three more event-accuracy points, while 60\% and 80\% calibration show smaller incremental gains. Within the memory protocol, larger calibration sets are associated with higher event and cascade accuracy. However, each row reports the strongest matcher from the evaluated candidate family for that ratio and seed, so the trend reflects both calibration-set size and matcher selection. The memory is evaluated separately and does not correct or specialize the pooled recognizer.

These calibration ratios are operating points, not a proposed product requirement. A deployed system should not ask users to label a fixed share of every future meal. In this study, the ratios compare controlled amounts of labeled target-user data. Table~\ref{tab:main} shows larger gains at lower ratios and smaller gains later. Because calibration time and user effort were not measured, this trend motivates, rather than establishes, a bounded setup or correction workflow.

\section{Discussion}

\noindent\textbf{From behavior labels to meal-awareness timelines.}
The central well-being value is not a single label, but a reviewable sequence that helps a user notice patterns. Oto-Meal is best suited to prompts such as ``several drinking-like segments were detected'' or ``this region is uncertain between drink and swallow,'' rather than normative dietary judgments. The system is therefore designed as a low-burden sensing substrate for reflection, not as a replacement for nutritional assessment. A behavior timeline can support lightweight questions about pace, hydration opportunities, conversation-heavy meals, or uncertain regions that deserve user correction. Future versions can expand the behavior vocabulary, but each added label should remain tied to user-understandable reflection rather than model-centric taxonomy growth.

\noindent\textbf{Robustness in realistic meal settings.}
The results also identify likely failure modes. Drink and swallow remain close, smaller-user estimates are more volatile, and real meals may add motion, speech, device placement changes, and temperature or physiological variation. These factors should be treated as design requirements for uncertainty displays and repeated-session calibration, not only as model errors. For example, a timeline interface could show lower confidence for drink/swallow regions, invite correction of representative segments, and prioritize collecting examples from meals that differ from the initial calibration. This keeps the limitations connected to concrete system design rather than presenting them as isolated weaknesses.

\noindent\textbf{Potential future directions.}
The current memory matcher is deliberately simple. Stronger future systems could combine the inspectability of calibration examples with learned PPG+IMU embeddings, self-supervised pretraining, style-aware adaptation, or distilled on-device models. The runtime result suggests that memory inference itself is not the bottleneck; the harder problem is collecting representative calibration examples and learning features that transfer across sessions. This creates a clear research path: keep the user-facing memory understandable, but improve the feature space that feeds it. Such a design would let future work evaluate stronger backbones while preserving the paper's main interaction principle: personalization should remain visible, correctable, and bounded.

\section{Conclusion}

In this paper, we propose Oto-Meal, which supports personalized meal-awareness review through audio- and image-free earable PPG+IMU sensing. By combining PPG+IMU windows, a two-stage behavior recognizer, and a user-conditioned memory, it supports reviewable meal-adjacent behavior timelines without raw audio, video, or food images. The seven-user evaluation shows that bounded target-user calibration can strengthen this sensing path beyond a pooled mixed-user model. Together, these results reveal the potential of pairing audio- and image-free earable sensing with bounded, inspectable calibration as a design path for low-burden and reflective meal-awareness tools.

\begin{acks}
This work was supported in part by the National Natural Science Foundation of China (NSFC) under Grant T2495254, in part by the SUSTech Fang Keng Faculty Award, and in part by the Center for Computing Science and Engineering at Southern University of Science and Technology. This work was also supported by the Special Funds for the Cultivation of Guangdong College Students' Scientific and Technological Innovation (``Climbing Program'' Special Funds), project no. pdjh2026c21104.
\end{acks}

\clearpage
\bibliographystyle{ACM-Reference-Format}
\bibliography{oto_meal_wellcomp2026_refs}

\end{document}